\documentclass[conference]{IEEEtran}

\usepackage{amsmath}
\usepackage{amssymb}
\usepackage{amsfonts}
\usepackage{bm}
\usepackage{graphicx}
\usepackage{cite}
\usepackage{color}
\usepackage{algorithm}
\usepackage{algorithmic}
\usepackage{multirow}
\usepackage{array}
\usepackage{booktabs}
\usepackage{url}
\usepackage{float}
\usepackage{subfigure}
\usepackage{stfloats}
\usepackage{float}
\usepackage{comment}
\usepackage{tikz}
\usepackage{cleveref}
\usepackage{acronym,balance}
\usetikzlibrary{arrows.meta, shapes.geometric, calc}

\allowdisplaybreaks

\newcommand{\eq}{=}
\newcommand{\urxi}{ \uu^{i}_{\rm{R}} }
\newcommand{\utx}{ \uu_{\rm{T}} }

\newcommand\norm[1]{\left\lVert #1\right\rVert}

\newcommand{\thn}[1]{ {#1^{\rm{th} } } }

\newcommand{\Tsym}{ T_{\rm{sym}} }

\newcommand{\Tcp}{ T_{\rm{cp}} }

\newcommand{\ptx}{ \pp_{\rm{T}} }
\newcommand{\prxi}{ \pp^{i}_{\rm{R}} }

\newcommand{\cc}{ \mathbf{c} }

\newcommand{\pp}{ \mathbf{p} }

\newcommand{\uu}{ \mathbf{u} }

\newcommand{\vv}{ \mathbf{v} }

\begin{document}
\bstctlcite{IEEEexample:BSTcontrol}
\acrodef{D-MIMO}{distributed MIMO}
\acrodef{SNR}{signal-to-noise ratio}
\acrodef{EFIM}{equivalent Fisher information matrix}
\acrodef{MMSE}{Minimum Mean Squared Error}

\acrodef{MSE}{mean square error}

\acrodef{PSD}{power spectral density}

\acrodef{RMSE}{root mean squared error}
\acrodef{SLR}{statistical linear regression}

\acrodef{AP}[AP]{access point}

\acrodef{ML}[ML]{maximum likelihood}
\acrodef{DBSCAN}[DBSCAN]{density-based spatial clustering of applications with noise}
\acrodef{UE}[UE]{user equipment}
\acrodef{BS}[BS]{base station}
\acrodef{VA}[VA]{virtual anchor}
\acrodef{VUE}[VUE]{virtual user equipment}
\acrodef{SP}[SP]{scattering  point}
\acrodef{IP}[IP]{incidence point}
\acrodef{fov}[FoV]{field-of-view}   
\acrodef{LoS}[LoS]{line-of-sight}
\acrodef{NLoS}[NLoS]{non-line-of-sight}
\acrodef{PMBM}[PMBM]{Poisson  multi-Bernoulli  mixture}
\acrodef{PMB(M)}[PMB(M)]{Poisson  multi-Bernoulli  (mixture)}
\acrodef{PMB}[PMB]{Poisson  multi-Bernoulli}
\acrodef{RFS}[RFS]{random finite set}
\acrodef{PPP}[PPP]{Poisson point process}
\acrodef{MBM}[MBM]{multi-Bernoulli  mixture}
\acrodef{MB}[MB]{multi-Bernoulli}

\acrodef{ekf}[EKF]{extended Kalman filter}
\acrodef{PDF}[PDF]{probability density function}

\acrodef{ckf}[CKF]{cubature Kalman filter}
\acrodef{rbp}[RBP]{Rao-Blackwellized particle}
\acrodef{gospa}[GOSPA]{generalized optimal subpattern assignment}
\acrodef{SLAM}[SLAM]{simultaneous localization and mapping}

\acrodef{TOA}[TOA]{time of arrival}
\acrodef{AOA}[AOA]{angles of arrival}
\acrodef{AOD}[AOD]{angles of departure}

\acrodef{IF}[IF]{information filter}
\acrodef{EIF}[EIF]{extended information filter}
\acrodef{kf}[KF]{Kalman filter}
\acrodef{PMF}[PMF]{probability mass function}
\acrodef{MAP}[MAP]{maximum a posteriori estimation}

\acrodef{DA}[DA]{data association}
\acrodef{OFDM}[OFDM]{Orthogonal Frequency Division Multiplexing}

\acrodef{PCRB}[PCRB]{posterior Cram{\'e}r-Rao bound}

\acrodef{EM}[EM]{expectation-maximization}
\acrodef{FOV}[FOV]{field of view}
\acrodef{RB}[RB]{Rao-Blackwellized }

\acrodef{ISAC}[ISAC]{integrated sensing and communication}

\acrodef{EK}[EK]{extended Kalman}
\acrodef{EKF}[EKF]{extended Kalman filter}

\acrodef{LMB}[LMB]{labeled multi-Bernoulli}
\acrodef{GLMB}[$\delta$-GLMB]{$\delta$-generalized labeled multi-Bernoulli}
\acrodef{PHD}[PHD]{probability hypothesis density}

\acrodef{OID}[OID]{optimal importance density}

\acrodef{MTT}[MTT]{multi-target tracking}

\acrodef{MCMC}[MCMC]{Markov chain Monte Carlo}

\acrodef{MH}[MH]{Metropolis-Hastings}

\acrodef{NMI}[NMI]{normalized mutual information}

\acrodef{CRB}[CRB]{Cram{\'e}r-Rao bound}

\acrodef{PIM}[PIM]{posterior information matrix}

\acrodef{OID}[OID]{optimal importance density}

\acrodef{MAE}[MAE]{mean absoluate error} 

\acrodef{FIM}[FIM]{Fisher information matrix}
\acrodef{PIM}[PIM]{posterior information matrix}
\acrodef{PEB}[PEB]{position error bound}
\acrodef{LEB}[LEB]{landmark error bound}
\acrodef{HEB}[HEB]{heading error bound}
\acrodef{CEB}[CEB]{clock bias error bound}

\acrodef{AoA}[AoA]{angles of arrival}
\acrodef{AoD}[AoD]{angles of departure}

\acrodef{URA}[URA]{uniform rectangular array}

\acrodef{RTT}{round-trip-time}

\acrodef{LS}{least-squares}

\acrodef{CDF}{cumulative distribution function}

\acrodef{AWGN}{additive white Gaussian noise}

\acrodef{MUSIC}{multiple signal classification}

\acrodef{S-MUSIC}{sequential multiple signal classification}
\acrodef{OMP}{orthogonal matching pursuit}
\acrodef{SCM}{spatial covariance matrix}

\acrodef{MF}{matched filtering}

\acrodef{AIC}{akaike information criterion}
\acrodef{MDL}{minimum description length}

\title{ Phase-Coherent Doppler-Only Sensing for 

D-MIMO ISAC}

\author{
    \IEEEauthorblockN{
        Silan Karadag\IEEEauthorrefmark{1}\IEEEauthorrefmark{2}, 
        Musa Furkan Keskin\IEEEauthorrefmark{1}, 
                Navid Amani\IEEEauthorrefmark{1}, 
        Nikita Lyamin\IEEEauthorrefmark{2},  
        Annelie Wyholt\IEEEauthorrefmark{2}, and
                Henk Wymeersch\IEEEauthorrefmark{1} 
\\
    }
    \IEEEauthorblockA{
        \IEEEauthorrefmark{1}Department of Electrical Engineering, Chalmers University of Technology, Gothenburg, Sweden
    }
    \IEEEauthorblockA{
        \IEEEauthorrefmark{2}Volvo Car Corporation, Gothenburg, Sweden
    }
    \IEEEauthorblockA{
        Email:         silan@chalmers.se, silan.karadag@volvocars.com
    }
}

\maketitle
\insert\footins{\noindent\footnotesize This research has been carried out in WiTECH Centre under the DisCourSe project financed by VINNOVA. This work is also supported, in part, by the Swedish Research Council (VR) through the project 6G-PERCEF under Grant 2024-04390.}
\begin{abstract}

This paper studies phase-coherent Doppler-only sensing for bandwidth-limited sub-6 GHz \ac{D-MIMO} \ac{ISAC}. We consider a multistatic system with one transmitter and multiple synchronized single-antenna receivers, and develop noncoherent and coherent maximum-likelihood estimators together with their position error bounds. The analysis shows that noncoherent sensing relies only on Doppler and cannot localize static targets, whereas phase-coherent processing exploits carrier phase to provide additional position information. Simulations confirm the bounds and show that coherent processing significantly improves localization accuracy in the considered narrowband setting.
\end{abstract}

\acresetall 
\section{Introduction}
\Ac{ISAC} has emerged as a key enabler of 6G, officially recognized by ITU-R in its IMT-2030 framework \cite{ITU_M2160} and currently under standardization by 3GPP \cite{3gpp_isac_tr22837}. Among the many envisioned applications, automotive sensing stands out as roadside units and distributed access points (APs) can assist onboard sensors for improved situational awareness, naturally giving rise to D-MIMO architectures that expand coverage beyond line-of-sight \cite{guo2025integrated}.

In sub-6 GHz bands, available bandwidth for sensing is limited due to regulatory constraints and competing spectrum demands \cite{Liu2022ISAC}, yet exploiting the spatial dimension through widely distributed antennas, inspired by multistatic radar principles, can overcome these bandwidth limitations \cite{MIMORadar}. Realizing this potential, however, hinges on the availability of phase coherence across the distributed nodes. Emerging C-RAN architectures \cite{Checko2015CloudRF} enable a shared oscillator across APs, thereby allowing for the joint processing of signals across spatially separated nodes which serves as the main enabler for cell-free MIMO. 

\begin{figure}
\centerline{\includegraphics[width=0.99\linewidth]{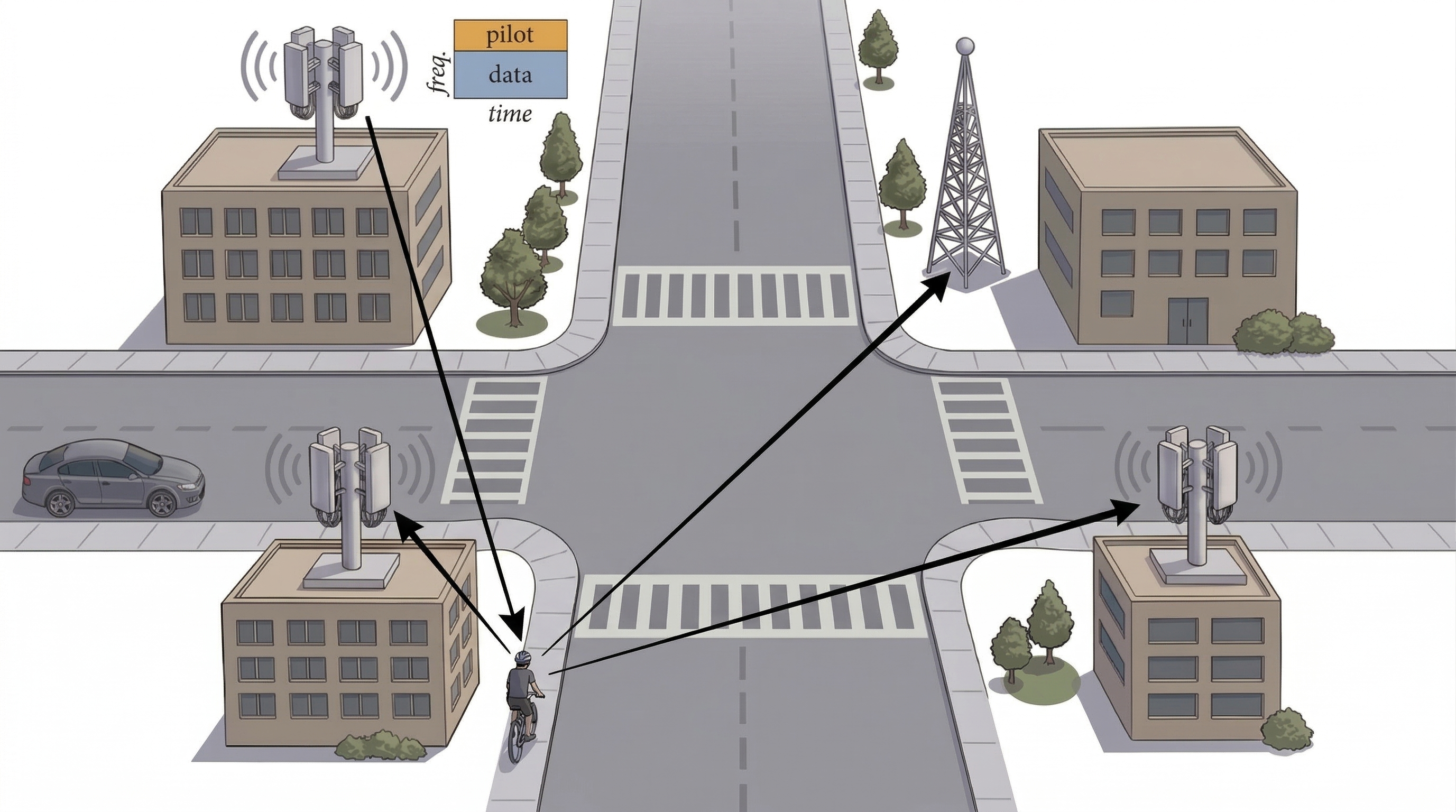}}
\caption{Infrastructure-assisted multistatic sensing in an urban intersection, based on Doppler from narrowband pilots.}
\label{fig:introduction}
\end{figure}

Early works in D-MIMO localization and sensing predominantly utilized non-coherent processing \cite{noncoherentradar}. As the demand for sub-meter accuracy grows, the transition to phase-coherent methods has become a critical research frontier. Interestingly, coherent methods have been mainly investigated 
for mmWave regimes, where large bandwidth inherently supports high-resolution range estimation \cite{s19204582, movISAC_2025}. In contrast, research in narrowband networks \cite{krysik2015} demonstrates that high precision localization can be achieved by leveraging Doppler measurements as a primary information source. Additionally, as shown in WiFi passive radar systems \cite{WIFIradar}, adding Doppler measurements to multistatic configurations improves moving target localization\cite{zhao_multistatic_doppler_2020}. Recent studies in the sub-6 GHz bands have demonstrated significant localization gains through carrier phase exploitation, though their focus remains on positioning rather than sensing  \cite{fascista2024joint, sauradeep}. Further ISAC studies investigated multi-target estimation for phase-coherent systems with multi-antenna APs \cite{venkatesh}. Despite these advancements, there remains a gap in the literature on Doppler-only sensing within phase-coherent ISAC frameworks for single-antenna nodes in bandwidth-constrained scenarios. 

To bridge the gap, this paper considers the problem of phase-coherent multistatic sensing based only on narrowband signals with distributed single-antenna \acp{AP} (see Fig.~\ref{fig:introduction}). The contributions are threefold: (i) we develop a single-subcarrier OFDM signal model for distributed sensing under the assumption of phase synchronization across APs;  (ii) we formulate both noncoherent and coherent maximum likelihood estimators for target localization in selected scenarios; (iii) To provide a performance benchmark and gain insight into the fundamental identifiability, we derive the corresponding noncoherent and coherent \acp{PEB}. To validate the bounds and estimators, we conduct  Monte-Carlo simulations.

\section{System Model}
We consider a D-MIMO architecture that is deployed with a single transmitter and $R$ receivers each equipped with a single antenna to demonstrate that sensing is possible with minimal resources while leveraging receiver diversity \cite{networkISAC2025}. The APs are synchronized in time, frequency, and phase, thus behaving as a distributed antenna array \cite{fascista2024joint}. 
\vspace{-0.05in}
\subsection{Geometry Model}
We consider a single isotropic point target. In a global 2D coordinate system, the transmitter AP and the $\thn{i}$ receiver AP have known and fixed positions denoted by $\ptx$ and $\prxi$, while the unknown location and velocity of the target are denoted by $\pp$ and $\vv$, respectively. 
\vspace{-0.05in}
\subsection{Transmit Signal Model}
Given our objective to utilize the Doppler effect as a primary sensing mechanism, we consider an OFDM waveform with $M$ slow-time symbols and a single active subcarrier. The transmitted symbols are represented by the vector $\mathbf{x} \in \mathbb{C}^{M \times 1}$, where $|x_m| = \sqrt{P_t}$. The total duration of a symbol is given by $\Tsym = \Tcp + T$, where $\Tcp$ and $T$ denote, respectively, the cyclic prefix (CP) and the elementary symbol durations. Since no wideband range estimation can be  performed, we refer to this as \textit{Doppler-only} sensing. 
\vspace{-0.05in}
\subsection{Geometric Channel Parameter Model}
\label{sec:channelParameters}
The bistatic delay $\tau_i$ for the $\thn{i}$ receiver is given by $
\tau_i = { \norm{\ptx - \pp}}/{c} + { \norm{\prxi - \pp} }/{c} $, where $\lambda = {c}/{f_c}$ denotes the  wavelength at the carrier $f_c$, and $c$ is the speed of light. The Doppler shift $\nu_i$ for the $\thn{i}$ receiver is given by $
\nu_i = {(\ptx - \pp)^{\top} \vv}/{(\norm{\ptx - \pp} \lambda)} + {(\prxi - \pp )^{\top} \vv}/{(\norm{\prxi - \pp} \lambda)} $. 
The magnitude of the combined channel gain of the two-segment path for the bistatic link associated with the $i$-th receiver
is denoted as $\alpha_i \in \mathbb{R}$, which involves the effect of both the bistatic radar cross section (RCS) and propagation-induced path loss. The phase induced by scattering off the isotropic target is denoted by $\vartheta$.  
\vspace{-0.05in}
\subsection{Received Signal Model}\label{sec_rec_sig_model}
We introduce the temporal steering vector $\mathbf{c}(\nu)\in  \mathbb{C}^M$ across $M$ slow-time symbols, indexed symmetrically as $m \in \{-(M-1)/2, \ldots, (M-1)/2\}$, by $[\mathbf{c}(\nu)]_m=\exp(j 2 \pi m T_{\mathrm{sym}}\nu)$. We introduce two observation models. 

\begin{itemize}
    \item 

\textbf{Coherent Operation:}
Under phase-coherence, the carrier phase   provides  information through the delay term $\tau_i$. The received signal at the $\thn{i}$ receiver can then be written as
\begin{align}
\label{eq:signal_model_y}
\mathbf{y}_i = \sqrt{P_t}\alpha_i \, e^{j \vartheta} e^{-j 2 \pi f_c \tau_i}  \mathbf{c}(\nu_i)  + \mathbf{z}_i,
\end{align}
where $\mathbf{z}_i \sim \mathcal{CN}(\mathbf{0}, \sigma_N^2 \mathbf{I})$ is the \ac{AWGN}, which is independent across all receivers. In \eqref{eq:signal_model_y}, the position $\pp$ can be inferred through both the delays $\tau_i$ and the Doppler shifts $\nu_i$.
\item 
\textbf{Noncoherent Operation:} The receivers can decide to ignore the phase coherence, in which case  the carrier phase can be absorbed into an unknown complex gain and the received signal in \eqref{eq:signal_model_y} can then be written as:
\begin{align}
\label{eq:signal_model_y_noncoh}
\mathbf{y}_i = \sqrt{P_t}\gamma_i \mathbf{c}(\nu_i) + \mathbf{z}_i,
\end{align}
where $\gamma_i = \alpha_i e^{j\vartheta} e^{-j2\pi f_c \tau_i}$ is an arbitrary unknown complex-valued parameter. 
Hence, in \eqref{eq:signal_model_y_noncoh}, $\pp$ can only be inferred through the Doppler shifts $\nu_i$.
\end{itemize}
\section{Fundamental Performance Bounds}
We first derive the \acp{PEB} for both observation models to establish the fundamental performance limits. 
We define the full parameter vector as $\boldsymbol{\theta} = [\boldsymbol{\xi}^\top, \boldsymbol{\eta}^\top]^\top$, where the parameters of interest are $\boldsymbol{\xi} = [\pp^\top, \vv^\top]^\top$, and $\boldsymbol{\eta}$ contains nuisance parameters that depend on the model. We seek to derive the \ac{EFIM} of the target position, denoted as $\mathbf{J}^{\text{C}}_{\mathbf{p}}$ (coherent case) and $\mathbf{J}^{\text{NC}}_{\mathbf{p}}$ (noncoherent case). Then, the \ac{PEB} is defined as  $\mathrm{PEB} = \sqrt{\rm{tr}(\mathbf{J}_{\mathbf{p}}^{-1} )}$.

\subsection{Noncoherent Model} 
\label{sec:3A1}
Given the signal model in \eqref{eq:signal_model_y_noncoh}, the nuisance parameters for the noncoherent case are $\boldsymbol{\eta} = [\Re(\gamma_1), \Im(\gamma_1),
...  , \Re(\gamma_R) , \Im(\gamma_R)]$. 
The FIM of $\boldsymbol{\theta}$ can be derived from the Slepian-Bangs formula \cite{kay1993statistical}
\begin{align}
 [\mathbf{J}^{\text{NC}}(\boldsymbol{\theta})]_{mn} = \frac{2}{\sigma_N^2} \sum_{i=1}^{R} \Re\!\left\{\frac{\partial \boldsymbol{\mu}_i^H}{\partial \theta_m} \frac{\partial \boldsymbol{\mu}_i}{\partial \theta_n} \right\},
\end{align}
where $\boldsymbol{\mu}_i$ is the noise-free signal. 
$\boldsymbol{\eta}$ and $\boldsymbol{\xi}$ become decoupled, leading to 
a block-diagonal FIM 
\begin{align}
\label{eq:block_diagonal_FIM_NCE}
  \mathbf{J}^{\text{NC}}(\boldsymbol{\theta}) =
  \begin{bmatrix}
    \mathbf{J}^{\text{NC}}_{\boldsymbol{\xi}\boldsymbol{\xi}} & \mathbf{J}^{\text{NC}}_{\boldsymbol{\xi}\boldsymbol{\eta}}  \\
    \mathbf{J}^{\text{NC}}_{\boldsymbol{\eta}\boldsymbol{\xi}}  & \mathbf{J}^{\text{NC}}_{\boldsymbol{\eta}\boldsymbol{\eta}}
  \end{bmatrix} =   \begin{bmatrix}
    \mathbf{J}^{\text{NC}}_{\boldsymbol{\xi}\boldsymbol{\xi}} & \mathbf{0} \\
    \mathbf{0} & \mathbf{J}^{\text{NC}}_{\boldsymbol{\eta}\boldsymbol{\eta}}
  \end{bmatrix}.
\end{align}
It can be readily shown that 
\begin{align}
\mathbf{J}^{\text{NC}}_{\boldsymbol{\xi}\boldsymbol{\xi}} =
  \mathbf{H}^\top \mathbf{W} \mathbf{H} 
  \label{eq:efim_4d_matrix}
\end{align}
where 
$\mathbf{W} = \mathrm{diag}(w_1, \dots, w_R)$ contains the per-link  Fisher information on the Doppler $\nu_i$, with $w_i = {2P_t |\alpha_i|^2}/{\sigma_N^2} \sum_{m} (2\pi m\,T_{\mathrm{sym}})^2$ 
and $\mathbf{H} \in \mathbb{R}^{R \times 4}$ is the Jacobian of the
Doppler shifts $\nu_i$ with respect to $\boldsymbol{\xi}$, with
$i$-th row $\mathbf{h}_i^\top =
[(\nabla_{\mathbf{p}}\nu_i)^\top,\, (\nabla_{\mathbf{v}}\nu_i)^\top]$. Introducing 
the unit direction vectors pointing from the target to the transmitter and the $i$-th receiver, respectively, as $\utx\eq{(\ptx - \pp)}/{\norm{\ptx - \pp}}$ and $ \urxi\eq{(\prxi - \pp)}/{\norm{\prxi - \pp}}$, and defining $\mathbf{d}_i \eq \utx\ +  \urxi $, we find
\begin{align}
  \nabla_{\mathbf{v}} \nu_i  & = \frac{1}{\lambda}
  \left( \utx + \urxi \right) = \frac{1}{\lambda}
  \mathbf{d}_i \,,
  \label{eq:grad_v}\\
    \nabla_{\mathbf{p}} \nu_i  & = - \frac{1}{\lambda}\boldsymbol{\Gamma}_i\mathbf{v} \,,\label{eq:grad_p}
\end{align}
where 
\begin{align}
    \boldsymbol{\Gamma}_i =  \frac{(\mathbf{I}_2 - \utx \utx^\top)}{\norm{\ptx - \pp}} +
  \frac{(\mathbf{I}_2 - \urxi (\urxi)^\top)}{\norm{\prxi - \pp}}.
\end{align}
Substituting \eqref{eq:grad_v} and \eqref{eq:grad_p} into
\eqref{eq:efim_4d_matrix}, the fully expanded 4D EFIM is given by
\begin{align}
  \mathbf{J}^{\text{NC}}_{\boldsymbol{\xi}\boldsymbol{\xi}} =
  \frac{1}{\lambda^2} 
  \begin{bmatrix}
    \sum_{i=1}^{R} w_i\boldsymbol{\Gamma}_i \mathbf{v}\mathbf{v}^\top \boldsymbol{\Gamma}_i &
    -\sum_{i=1}^{R} w_i\boldsymbol{\Gamma}_i \mathbf{v}\mathbf{d}_i^\top \\
    -\sum_{i=1}^{R} w_i\mathbf{d}_i \mathbf{v}^\top \boldsymbol{\Gamma}_i &
    \sum_{i=1}^{R} w_i\mathbf{d}_i \mathbf{d}_i^\top
  \end{bmatrix}, 
  \label{eq:efim_4d_expanded}
\end{align}
so that 
\begin{align} 
    \mathbf{J}^{\text{NC}}_{\mathbf{p}} & = \frac{1}{\lambda^2} \sum_{i=1}^{R} w_i  \boldsymbol{\Gamma}_i \mathbf{v}\mathbf{v}^\top \boldsymbol{\Gamma}_i \label{eq:EFIMNC} \\      
     -\frac{1}{\lambda^2}&\Big( \sum_{i=1}^{R} w_i\boldsymbol{\Gamma}_i \mathbf{v}\mathbf{d}_i^\top\Big)\Big(\sum_{i=1}^{R} w_i\mathbf{d}_i \mathbf{d}_i^\top\Big)^{-1}\Big(\sum_{i=1}^{R} w_i\mathbf{d}_i \mathbf{v}^\top \boldsymbol{\Gamma}_i\Big). \notag 
\end{align}

From \eqref{eq:EFIMNC}, we deduce that the  information on the position scales with target speed and decays with distance through $\boldsymbol{\Gamma}_i$. When the target is static, the position is not identifiable. 
Moreover, $\sum_{i=1}^{R} w_i\mathbf{d}_i \mathbf{d}_i^\top$ must be full-rank (hence $R\ge 2$) to avoid catastrophic degradation due to unknown target velocity. This suggests that targets that are collinear with all the receivers and do not lie between any of them cannot be localized.

\subsection{Coherent Model}
\label{sec:3A2}
Given the signal model in \eqref{eq:signal_model_y}, the nuisance parameters for the coherent case are 
$\boldsymbol{\eta} = [\alpha_1, \ldots, \alpha_R, \vartheta]^\top$. Defining $ \phi_i \eq \vartheta - 2\pi f_c \tau_i $, $\boldsymbol{\mu}_i = {\alpha_i e^{j\phi_i}}{\mathbf{c}(\nu_i)}$. In this case, the nuisance parameters are no longer decoupled from the target state: 
\begin{align}
     \mathbf J^{\text{C}}_{\boldsymbol{\xi} \boldsymbol{\eta}}
    = \begin{bmatrix}
        \mathbf 0_{2\times R} & \mathbf J^{\text{C}}_{\mathbf p\vartheta} \\
        \mathbf 0_{2\times R} & \mathbf 0_{2\times 1}
    \end{bmatrix} \neq \mathbf{0}.
\end{align}
where $  \mathbf J^{\text{C}}_{\mathbf p\vartheta}
    = {4\pi P_t M}/{(\sigma_N^2\lambda)} \times
      \sum_{i=1}^R \alpha_i^2 \mathbf d_i$. After some manipulations, we  find that 
\begin{align}
\mathbf{J}^{\text{C}}_{\boldsymbol{\xi}\boldsymbol{\xi}} & =\mathbf{J}^{\text{NC}}_{\boldsymbol{\xi}\boldsymbol{\xi}}
   + \begin{bmatrix}        
        \mathbf{J}_{\text{cp}}& \mathbf 0 \\
        \mathbf 0 & \mathbf 0
    \end{bmatrix}
\end{align}
where $\mathbf{J}_{\text{cp}}$ comprises two terms: a positive information term that captures the additional information on the target position thanks to the carrier information, and a negative information term (i.e., information loss) due to $\vartheta$ being unknown: 
\begin{align}
    \mathbf{J}_{\text{cp}} &  = \frac{8\pi^2 P_t M}{\sigma_N^2\lambda^2} \bigg(\sum_i \alpha_i^2\mathbf d_i\mathbf d_i^\top -  
          \frac{
            \left(\sum_i \alpha_i^2\mathbf d_i\right)
            \left(\sum_i \alpha_i^2\mathbf d_i\right)^\top
          }{\sum_i \alpha_i^2}\bigg)\notag \\
          & = 
\frac{8\pi^2P_t M}{\sigma_N^2\lambda^2}
\sum_i \alpha_i^2(\mathbf d_i-\bar{\mathbf d})(\mathbf d_i-\bar{\mathbf d})^\top,
\end{align}
where $\bar{\mathbf d}
=
{\sum_i \alpha_i^2\mathbf d_i}/{\sum_i \alpha_i^2}$. 
Hence, the coherent gain $\mathbf{J}_{\text{cp}} \succeq \mathbf{0}$ depends on the weighted spread of the bistatic direction vectors $\mathbf d_i$. If all $\mathbf d_i$ are identical or nearly aligned, phase coherence brings little or no extra position information. If the $\mathbf d_i$ are diverse, the coherent gain is large.
 It then immediately follows that 
\begin{align}
\mathbf{J}^{\text{C}}_{\mathbf{p}} = \mathbf{J}^{\text{NC}}_{\mathbf{p}} + \mathbf{J}_{\text{cp}} \succeq \mathbf{J}^{\text{NC}}_{\mathbf{p}}.
\end{align}
From these derivations, we can draw several conclusions: phase coherence can only improve the information on the target position, though the bistatic geometry plays an important role. Moreover, phase coherence allows us to estimate static targets under sufficiently diverse bistatic geometry, which was impossible under noncoherent operation. 

\section{Estimators}

We now design two estimators: (i) noncoherent estimator (NCE) that does not use the carrier phase and (ii) coherent estimator (CE) that exploits the carrier phase information. In both cases, we rely on the \ac{ML} criterion: 
\begin{align}
\hat{\boldsymbol{\theta}} & = \arg \max_{\boldsymbol{\theta}} \ln p(\mathbf{y} | \boldsymbol{\theta})= \arg \min_{\boldsymbol{\theta}} \mathcal{L}(\boldsymbol{\theta}),
\end{align} where 
$\mathcal{L}(\boldsymbol{\theta})
  = \sum_{i=1}^{R}
    \lVert \mathbf{y}_i - \boldsymbol{\mu}_i(\boldsymbol{\theta}) \rVert^2$.
\subsection{Noncoherent Estimator (NCE)}
In the noncoherent model~\eqref{eq:signal_model_y_noncoh}, the nuisance at the $i$-th receiver is the unknown complex gain $\gamma_i$. Minimizing $\mathcal{L}_i=\lVert \mathbf{y}_i - \sqrt{P_t }\gamma_i \mathbf{c}(\nu_i) \rVert^2$ with respect to $\gamma_i$ yields the least-squares estimate
$\hat{\gamma}_i = {\mathbf{c}^H(\nu_i) \mathbf{y}_i}/{(M\sqrt{P_t })}$.
Substituting $\hat{\gamma}_i$ back into $\sum_i\mathcal{L}_i$ gives the noncoherent cost function:
\begin{equation}
    \mathcal{L}_{\mathrm{NCE}}(\mathbf{p},\mathbf{v}) =  - \frac{1}{M} \sum_{i=1}^{R}\bigl| \mathbf{y}_i^H \mathbf{c}(\nu_i(\mathbf{p},\mathbf{v})) \bigr|^2.
    \label{eq:L_NCE_cost}
\end{equation}
The position and velocity estimate is then obtained by:
\begin{align}
    (\hat{\mathbf{p}}, \hat{\mathbf{v}})_{\mathrm{NCE}} = \arg \max_{(\mathbf{p},\mathbf{v})}  \sum_{i=1}^{R} {| \mathbf{y}_i^H \mathbf{c}(\nu_i(\mathbf{p},\mathbf{v})) |^2} \label{eq:L_NCE_argmax}
\end{align}
where $\nu_i(\mathbf{p},\mathbf{v})$ was introduced in Sec.~\ref{sec_rec_sig_model}.

\subsection{Coherent Estimator (CE)}

We introduce $\mathbf{s}_i(\mathbf{p},\mathbf{v}) \triangleq \sqrt{P_t }\, e^{-j2\pi f_c\tau_i(\mathbf{p})}\mathbf{c}(\nu_i(\mathbf{p},\mathbf{v}))$, so that the ML cost becomes $
\mathcal{L}(\boldsymbol{\theta})
= \sum_{i=1}^{R}
\lVert \mathbf{y}_i - \alpha_i\, e^{j\vartheta}\, \mathbf{s}_i(\mathbf{p},\mathbf{v}) \rVert^2.
\label{eq:L_theta_coh} $
Minimizing $\mathcal{L}(\boldsymbol{\theta})$ with respect to $\alpha_i$ yields
$
\hat{\alpha}_i(\mathbf{p},\mathbf{v},\vartheta)
= {\Re\bigl\{e^{-j\vartheta}\,\mathbf{s}_i^H(\mathbf{p},\mathbf{v})\,\mathbf{y}_i\bigr\}}/
       {\lVert \mathbf{s}_i(\mathbf{p},\mathbf{v}) \rVert^2}.
\label{eq:alpha_hat_CE}
$
Substituting $\alpha_i$ back 
gives the coherent cost function: $
\mathcal{L}(\mathbf{p},\mathbf{v},\vartheta)
= 
- \sum_{i=1}^{R}
{\bigl(\Re\bigl\{e^{-j\vartheta}\,\mathbf{s}_i^H(\mathbf{p},\mathbf{v})\,\mathbf{y}_i\bigr\}\bigr)^2}/
     {\lVert \mathbf{s}_i(\mathbf{p},\mathbf{v}) \rVert^2}.
\label{eq:L_sub_CE} $
Since $\lVert \mathbf{s}_i(\mathbf{p},\mathbf{v}) \rVert^2 = P_t M$, we define $
\rho_i(\mathbf{p},\mathbf{v})
\triangleq \mathbf{s}_i^H(\mathbf{p},\mathbf{v})\,\mathbf{y}_i.
\label{eq:rho_def} $
Using the identity $(\Re\{z\})^2 = \tfrac{1}{2}(|z|^2 + \Re\{z^2\})$, we obtain
\begin{align}
\mathcal{L}(\mathbf{p},\mathbf{v},\vartheta)
= 
- \frac{1}{2P_t M}
\Biggl[
  \sum_{i=1}^{R} |\rho_i|^2
  + \Re\!\biggl\{e^{-j2\vartheta}\sum_{i=1}^{R} \rho_i^2\biggr\}
\Biggr].
\label{eq:L_CE_rho}
\end{align}
Hence, the optimal phase can be written as $
\hat{\vartheta}(\mathbf{p},\mathbf{v})
= \frac{1}{2}\,\angle\!\big(\sum_{i=1}^{R} \rho_i^2(\mathbf{p},\mathbf{v})\big).
\label{eq:theta_hat_CE} $
Substituting $\hat{\vartheta}$ back yields the concentrated cost function
\begin{align}
\mathcal{L}_{\mathrm{CE}}(\mathbf{p},\mathbf{v})
&= 
- \frac{1}{2P_t M}
\biggl[
  \sum_{i=1}^{R} |\rho_i|^2
  + \Bigl|\sum_{i=1}^{R} \rho_i^2\Bigr|
\biggr].
\label{eq:L_CE_cost}
\end{align}
Dropping irrelevant constants, the cost function simplifies to
\begin{align}
\mathcal{L}_{\mathrm{CE}}(\mathbf{p},\mathbf{v})
&= - \biggl[
  \sum_{i=1}^{R} |\rho_i(\mathbf{p},\mathbf{v})|^2
  + \Bigl|\sum_{i=1}^{R} \rho_i^2(\mathbf{p},\mathbf{v})\Bigr|
\biggr].
\label{eq:L_CE_cost}
\end{align}
Similar to \eqref{eq:L_NCE_argmax}, the coherent ML estimator is
\begin{align}
(\hat{\mathbf{p}},\hat{\mathbf{v}})_{\mathrm{CE}}
= \arg\max_{\mathbf{p},\mathbf{v}}
  \sum_{i=1}^{R} |\rho_i(\mathbf{p},\mathbf{v})|^2
  + \biggl|\sum_{i=1}^{R} \rho_i^2(\mathbf{p},\mathbf{v})\biggr|.
\label{eq:L_CE_argmax}
\end{align}
We note that \eqref{eq:L_CE_argmax} breaks down into 2 terms: the non-coherent term from \eqref{eq:L_NCE_argmax} and an additional coherent term. 

\subsection{Solving the Optimization Problems}
Solving the optimization problems  \eqref{eq:L_NCE_argmax} and \eqref{eq:L_CE_argmax} can be achieved via gradient descent based on an initial estimate (from prior knowledge such as road layout) or by
alternating optimization between $\mathbf{p}$ and $\mathbf{v}$. 

\section{Results \& Discussion}

\subsection{Simulation Scenario and Parameters}

The simulation environment consists of a $50 \times 50$ m area with a transmitter at the origin and seven receivers distributed to provide high spatial diversity, as illustrated in Fig.~\ref{fig:cost_analysis}(b). A point target with a randomized scattering phase $\vartheta$ is placed at $\mathbf{p} = [20, 15]^{\top}$\,m with velocity $\mathbf{v} = [-5, 1]^{\top}$\,m/s. All APs are assumed to be equipped with a single antenna and share a common local oscillator, which enables phase-coherent signal processing. 
The total noise power $\sigma_N^2$ is determined by the thermal noise floor, the receiver noise figure, and the system bandwidth:
$\sigma_N^2 = k_B T_n F B$ where $k_B = 1.38 \times 10^{-23}$ J/K is the Boltzmann constant, $T_n = 290$ K is the standard noise temperature, $B$ is the system bandwidth in Hz, and $F$ is the noise figure in linear scale. The bistatic radar equation in \cite{richards2005fundamentals} gives
$    \alpha_i = \sqrt{(G_t G_r \sigma \lambda^2)/((4\pi)^3 R_{t}^2 R_{r_i}^2)} $
where $G_t$ and $G_r$ represent the antenna gains of the transmitter and the $i$-th receiver, respectively, $\sigma$ is the radar cross-section (RCS) of the target. The terms $R_t = \norm{\ptx - \pp}$ and $R_{r_i} = \norm{\prxi - \pp}$ denote the distances from the transmitter to the target and from the target to the $i$-th receiver, respectively.
%
%
%
Parameters follow the default setup in Table~\ref{tab:params}, except for those varied in the simulation results.

\begin{table}
\centering
\caption{Simulation Parameters}
\label{tab:params}
\begin{tabular}{llll}
\toprule
\textbf{Parameter} & \textbf{Symbol} & \textbf{Value} & \textbf{Unit} \\
\midrule
Carrier frequency & $f_c$ & $3.5$ & GHz \\
Subcarrier Spacing & $\Delta f$ & $60$ & kHz \\
Number of subcarriers & $N$ & $1$ & -- \\
Number of symbols & $M$ & $127$ & -- \\
Bandwidth & $B$ & $60$ & kHz \\
Noise Figure & $F$ & $8$ & dB \\
Transmitter Gain & $G_t$ & $0$ & dBi \\
Receiver Gain & $G_r$ & $0$ & dBi \\
Target RCS & $\sigma$ & $1$ & m$^2$ \\
Transmit Power & $P_t$ & $10$ & dBm \\
\bottomrule
\end{tabular}
\end{table}

\vspace{-0.1in}
\subsection{Cost Function Analysis}

\begin{figure*}[!t]
    \centering
\includegraphics[width=0.9\textwidth]{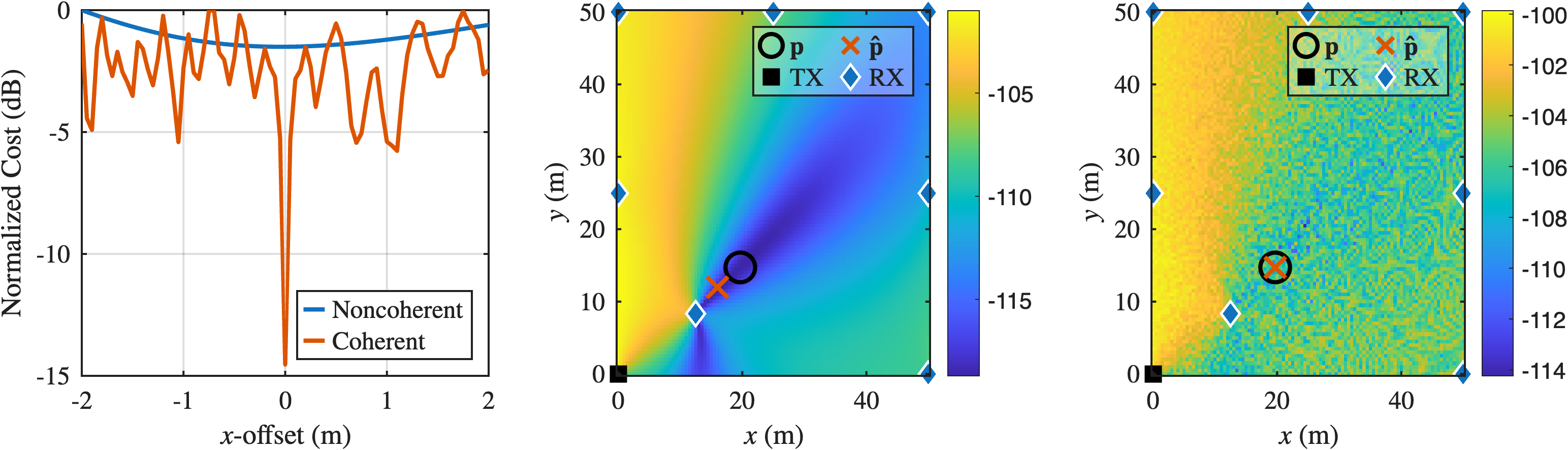}    
    \begin{small}
    \begin{tabular}{p{0.3\linewidth} p{0.3\textwidth} p{0.3\textwidth}}
        \centering (a) & \centering (b) & \centering (c)
    \end{tabular}
    \end{small}
    \caption{Cost function analysis: (a) 1D cost around true position; (b) 2D map of $ \mathcal{L}_{\mathrm{NCE}}$; (c) 2D map of $ \mathcal{L}_{\mathrm{CE}}$.}
    \label{fig:cost_analysis}\vspace{-5mm}
\end{figure*}

To gain insight into the estimation, we examine the structure of the cost functions in \eqref{eq:L_NCE_cost} and \eqref{eq:L_CE_cost} through 1D cuts and 2D spatial maps, at the true target velocity, which simplifies the visualization to a 2D spatial domain. As seen in Fig.~\ref{fig:cost_analysis}(a), the 1D cut along the $x$-axis through the true target position reveals a fundamental trade-off: the coherent cost function produces a deep, narrow minimum at the true location, whereas the noncoherent cost function remains shallow. The 2D maps tell the same story, i.e., the noncoherent map in Fig.~\ref{fig:cost_analysis}(b) is smooth, while the coherent map in Fig.~\ref{fig:cost_analysis}(c) is densely oscillatory due to phase variations in $e^{-j2\pi f_c \tau_i}$, creating many local minima across the search space.

\begin{figure*}[!t]
    \centering
\includegraphics[width=0.9\textwidth]{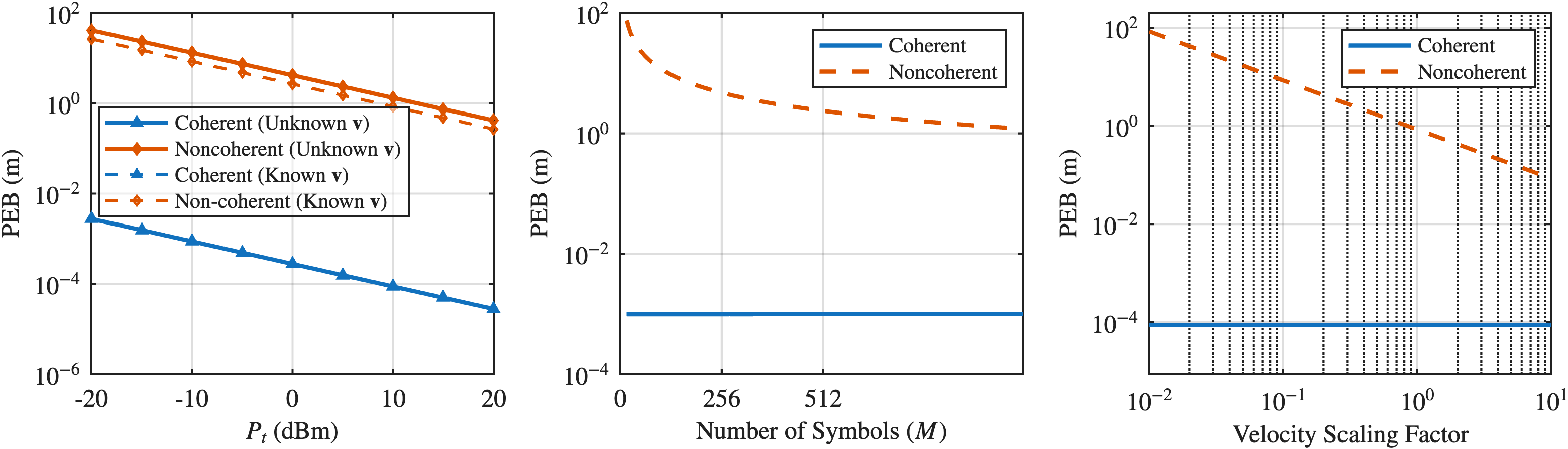}    
    \begin{small}
    \begin{tabular}{p{0.3\linewidth} p{0.3\textwidth} p{0.3\textwidth}}
    \centering (a) & \centering (b) & \centering (c)
    \end{tabular}
    \end{small}
    \caption{PEB analysis for: (a) known and unknown velocity; (b) number of symbols; (c) target mobility (scaled relative to $\mathbf{v} = [-5, 1]^\top$ m/s).}
    \label{fig:peb_rmse_plots}\vspace{-5mm}
\end{figure*}

\vspace{-0.1in}
\subsection{PEB Analysis}
The following analysis examines the PEB to characterize the impact of key system parameters.
\subsubsection{Impact of Known vs.~Unknown $\mathbf{v}$}
Fig.~\ref{fig:peb_rmse_plots}a shows the impact of velocity uncertainty on localization. The noncoherent PEB suffers a slight performance gap between the known and unknown velocity cases, as errors in velocity estimation propagate directly into the position estimate. The coherent PEB, by contrast, shows nearly identical performance in both cases, since the carrier phase $e^{-j2\pi f_c \tau_i}$ provides the dominant geometric constraint on position that is independent of Doppler. This further highlights the robustness of phase-coherent processing in dynamic environments where target velocity may not be known a priori.

\subsubsection{Impact of $M$}
Fig.~\ref{fig:peb_rmse_plots}b shows the PEBs as a function of the number of symbols $M$, keeping the total transmit energy fixed ($P_t M T_{\rm sym}$). The noncoherent PEB decreases significantly with increasing $M$ as the FIM in \eqref{eq:EFIMNC} scales with $M$ through $w_i$, described after \eqref{eq:efim_4d_matrix} (notice that $w_i$ scales with $P M^2$). As we collect samples over a longer time duration (and, thus enlarge the time aperture in $\cc(\nu)$), more Doppler information is accumulated, leading to better positioning. In contrast, the coherent PEB stays almost constant with respect to $M$ since the dominant source of information for positioning comes from the delay-dependent carrier phase term in \eqref{eq:signal_model_y} and the Doppler-related information plays a negligible role.



\subsubsection{Impact of $\mathbf{v}$}
Fig.~\ref{fig:peb_rmse_plots}c shows the PEBs versus target velocity where $P_t = 10$ dBm. The coherent bound remains flat, as carrier phase dominates its precision. In contrast, the noncoherent bound relies entirely on Doppler shifts, failing for stationary targets but improving significantly with speed in compliance with the analysis after \eqref{eq:EFIMNC}.
\vspace{-0.1in}
\subsection{Estimator Performance}
\begin{figure}
    \centering
\centerline{\includegraphics[width=0.9\linewidth]{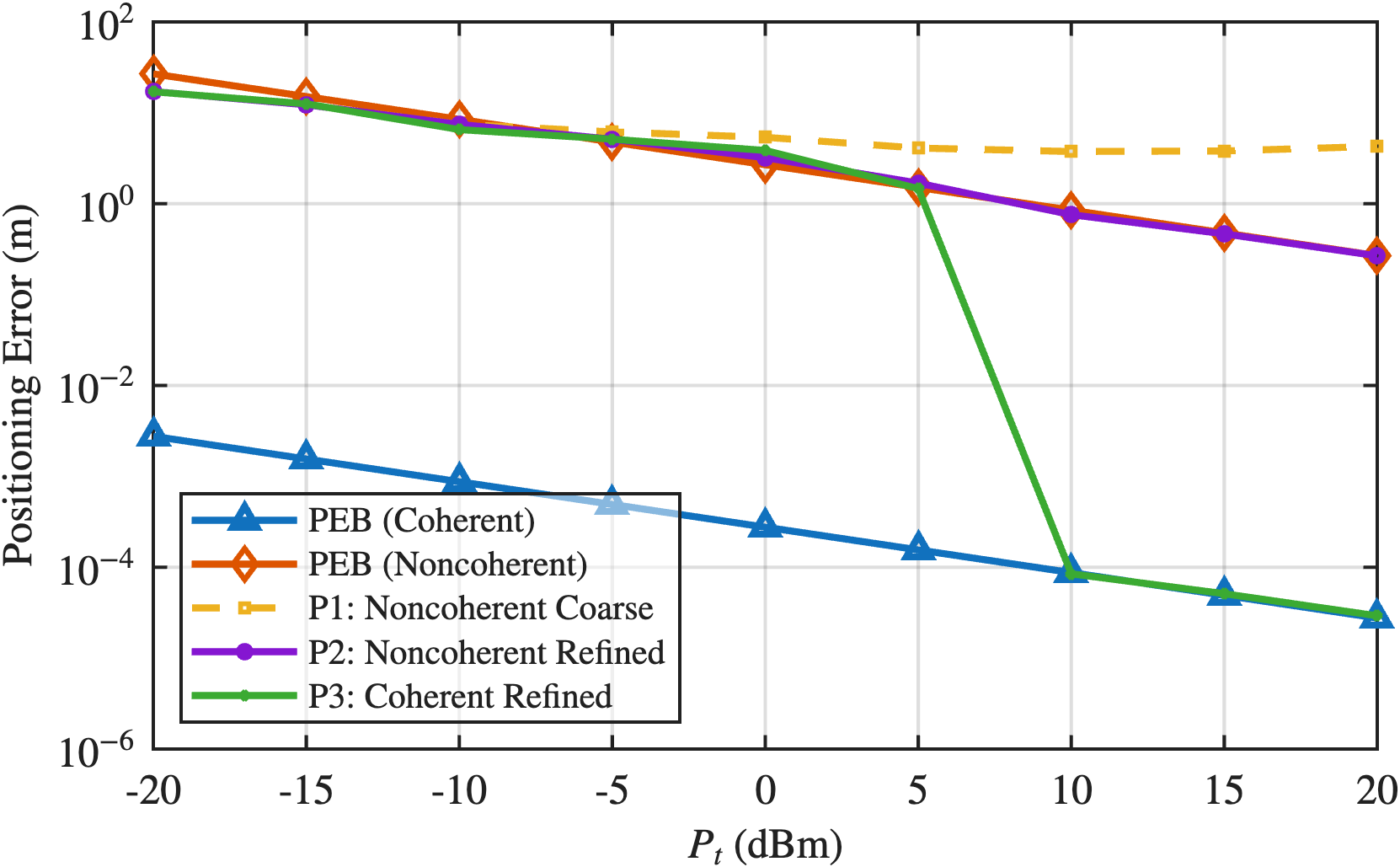}}    
    \caption{Transmit power vs. position RMSE.}
    \label{fig:TX_snr}
\end{figure}

To understand if the PEB is attainable, we developed a simple estimator. For this preliminary study, we consider the velocity to be known from other sensors. In light of the cost function behavior observed in Fig.~\ref{fig:cost_analysis}, we implement a two-stage approach: a global grid search using $\mathcal{L}_{\text{NCE}}$ to identify the coarse location of the target, denoted as P1, followed by a local grid search using $\mathcal{L}_{\text{CE}}$ and a gradient-based refinement using $\mathcal{L}_{\text{CE}}$, denoted as P3. We also define P2 as P1 followed by a gradient-based refinement using $\mathcal{L}_{\text{NCE}}$. This way we retain the precision of phase coherence while avoiding sidelobe ambiguities, and the computational overhead is significantly reduced. 
We evaluate localization performance via Monte Carlo simulations and compare the estimators against the  PEBs. Fig.~\ref{fig:TX_snr} shows the position root mean squared error (RMSE) as a function of transmit power. While all estimators asymptotically converge toward their respective bounds, the initial coarse grid search (P1) is limited by the grid resolution therefore preventing it from reaching theoretical limit. However, by employing a subsequent gradient-based refinement (P2) we can achieve the PEB. The coherent estimator (P3) shows a threshold effect, where sidelobe ambiguities arising at low SNR due to noise-induced ghost peaks are resolved with increasing transmit power, allowing the estimator to attain the PEB. 

\section{Conclusion}
This paper studied phase-coherent Doppler-only sensing for bandwidth-limited sub-6 GHz D-MIMO ISAC. We considered a multistatic setup with one transmitter and multiple synchronized single-antenna receivers, and developed both noncoherent and coherent maximum-likelihood estimators together with their corresponding position error bounds. The results show that noncoherent sensing relies entirely on Doppler information and therefore cannot localize static targets, whereas phase-coherent processing exploits carrier phase to provide additional geometric information. As a result, coherent processing achieves clear localization gains in the considered narrowband setting, which is also confirmed by the simulation results.
Several directions remain for future work: implementation of joint position-velocity estimation, assessing robustness to practical impairments (synchronization and calibration errors across distributed APs), and extending the framework to  more realistic scenarios with 3D geometries, multiple targets, clutter, and multipath.
\vspace{-0.15in}

\balance
\bibliographystyle{IEEEtran}
\bibliography{isac}    

\end{document}